\documentclass[amsmath,amssymb,aps,eqsecnum,nofootinbib,preprintnumbers]{revtex4}
\usepackage[dvips]{graphics,color}
\usepackage{latexsym}
\usepackage{graphicx}

\begin{document}

\preprint{KUNS-2113}

\title{Orthogonal black di-ring solution}

\author{
Keisuke Izumi\footnote{e-mail:
ksuke@tap.scphys.kyoto-u.ac.jp}
\\~}

~\\

\address{Department of Physics, Kyoto University, Kyoto 606-8502, Japan}

\begin{abstract}
We construct a five dimensional exact solution of the orthogonal black di-ring
which has two black rings  
whose $S^1$-rotating planes are orthogonal.
This solution has four free parameters 
which represent 
radii of 
and speeds of $S^1$-rotation of the black rings. 
We use the inverse scattering method.
This method needs the seed metric.
We also present a systematic method how to construct a seed metric. 
Using this method, we can probably construct other solutions having
many black rings on the two orthogonal planes with or 
without a black hole at the center. 
\end{abstract}
\maketitle

\section{Introduction}

In recent years, higher dimensional black objects
have been actively studied.
This is partly because string theory tells us that 
the spacetime we live in is higher dimensional.
In order for us to feel that the number of spacetime dimensions is four on large scales, 
the extra dimensions must be compactified.
However, when we observe a phenomenon on a small scale, 
the effect of higher dimensions may appear.
If the compactification scale is sufficiently large,
this effect can be detected in future collider experiments~\cite{collider}.

The only asymptotically flat static solution of the vacuum Einstein equations 
in higher dimensions is Schwarzschild-Tangherlini solution~\cite{Tan,static}, 
which is stable against perturbation~\cite{Ishibashi}.
These properties are the same as in the four dimensional case.
However, in the asymptotically flat, 
stationary and axisymmetric case,
the uniqueness theorem of the black hole 
does not exist in higher dimensions unlike the four dimensional case.
Myers and Perry discovered the higher dimensional black hole~\cite{Myers}
whose topology is $S^{D-1}$, 
which is an extension of the Kerr black hole to higher dimensions.
This solution was obtained also by the solitonic solution-generating methods~\cite{Iguchi2} 
and the inverse scattering method \cite{Tomizawa3}.
A five dimensional Myers-Perry black hole is unique~\cite{Morisawa} 
if black hole topology is $S^3$ in the asymptotically flat spacetime 
and if the spacetime has three commuting Killing vectors.
The black ring solution with horizon topology $S^1 \times S^2$
was discovered by Emparan and Reall~\cite{Emparan}.
The black ring rotating on the $S^1$ plane, 
which helps the balance against its attractive self-gravity force.
The $S^2$-rotating black ring solution was 
discovered by Mishima and Iguchi~\cite{Mishima,Tomizawa2,Figueras},
in which a plane supporting black ring from falling due to the self-gravity
is needed.
The solution of black ring with $S^1$ and $S^2$ rotations was also found
by Pomeransky and Sen'kov~\cite{Pomeransky}, 
which first had been found by numerical method~\cite{Kudoh}. 
It was proved that the black ring solution with horizon topology $S^1 \times S^2$
is only  Pomeransky and Sen'kov solution~\cite{Yazadjiev,Tomizawa}.
In the case with Maxwell fields, a generalzed analysis has done~\cite{Yazadjiev6}.
Moreover, the solution which has a number of black objects, 
such as black di-ring~\cite{diring,diring2} and black saturn~\cite{saturn} 
were also discovered. 
Rotating dipole black ring solution~\cite{Yazadjiev2,Yazadjiev3,Yazadjiev5} 
 and black saturn with dipole black ring solution~\cite{Yazadjiev4} have been generated in five-dimensional 
Einstein-Maxwell-dilaton gravity.

In this paper, we construct the solution having 
two black rings which are orthogonal to each other.
We call this solution ``orthogonal black di-ring".
We use the inverse scattering method~\cite{ISM}.
In inverse scattering method, 
we need a seed solution.
We present a method of constructing a diagonal seed metric.

This paper is organized as follows.
In Sec.~\ref{inverse}, we will review the inverse scattering method.
In Sec.~\ref{seed}, we will present a method of constructing a seed metric, 
giving a seed metric for the orthogonal black di-ring solution. 
In Sec.~\ref{trans}, we will show how to obtain the
orthogonal black di-ring solution using the inverse scattering method 
and we will write the obtained metric explicitly.
In Sec.~\ref{analysis}, we will analyse the regularity of the obtained solution.
The solution with several parameters generally has singularities.
However, we can remove all of these singularities 
if we choose the  parameters appropriately, 
leaving four free parameters.
We show the conditions that the parameters must satisfy for regularity.
In Sec.\ref{summary}, we will summarize our results.

\section{Inverse scattering method} \label{inverse}

In this section, we briefly explain the inverse scattering method~\cite{ISM}, 
by which a new metric can be obtained from a known seed metric.
This method can be used when the metric which we want has
$D-2$ commuting killing vector fields, one of which is timelike,
 in $D$-dimensional spacetime.
In this paper, we only present the procedure for generating a new solution
without giving its deviation.
For detailed deviation, see Belinsky's Paper~\cite{ISM}.

Thanks to the symmetries, the metric can  be written as
\begin{eqnarray}
ds^2 = f (d\rho^2+ dz^2) + g_{ab} dx^a dx^b, \label{metric}
\end{eqnarray}
where $f$ and $g_{ab}$ depend only on $\rho$ and $z$.
Here without loss of generality, 
we can set the determinant of $g_{ab}$ as
\begin{eqnarray}
\det g_{ab} =-\rho^2.
\end{eqnarray}
Then the Einstein equations become
\begin{eqnarray}
&&\partial_\rho U + \partial_z V =0, \label{diffg}\\
&&U=\rho(\partial _\rho {\bf g}) {\bf g}^{-1},
\qquad
V=\rho(\partial _z {\bf g}) {\bf g}^{-1},
\label{UV}\\
&&\partial_\rho \ln f = 
-\frac{1}{\rho} +\frac{1}{4\rho} \mbox{Tr}(U^2-V^2),
\label{rhof}\\
&&\partial_z \ln f = \frac{1}{2\rho} \mbox{Tr}(UV),\label{zf}
\end{eqnarray}
where ${\bf g}$ is $g_{ab}$ in matrix notation.
These equations can be classified.
The first three are the differential equations of ${\bf g}$.
The others are the equations from which $f$ is obtained for a given ${\bf g}$. 
From eq.(\ref{diffg}), 
we find that integrability condition for $\ln f$, 
$\partial_\rho\partial_z \ln f = \partial_z\partial_\rho \ln f$,
is satisfied.

Suppose that a seed metric ${\bf g_0}$
which satisfies the Einstein equations (\ref{diffg}-\ref{zf}) is prepared.
Then we consider linear differential equations
\begin{eqnarray}
&&\left(\partial_z 
-\frac{2 \lambda^2}{\lambda^2+\rho^2}\partial_\lambda\right) \Psi
 =\frac{\rho V_0-\lambda U_0}{\lambda^2 +\rho^2} \Psi,
 \label{psi1}
 \\
&&\left( \partial_\rho +
\frac{2 \lambda \rho}{\lambda^2+\rho^2}\partial_\lambda \right) \Psi
 =\frac{\rho U_0-\lambda V_0}{\lambda^2 +\rho^2} \Psi, 
 \label{psi2}
\end{eqnarray}
where $U_0$ and $V_0$ are $U$ and $V$ with ${\bf g}={\bf g_0}$, 
$\lambda$ is the complex spectral parameter 
independent of $\rho$ and $z$, 
and 
$\Psi=\Psi(\lambda,\rho,z)$ is a $(D-2)\times(D-2)$ matrix.
Solving these equations, we get the matrix $\Psi_{g_0}$.

We prepare functions $\mu_k(\rho,z)$ and $\bar \mu_k(\rho,z)$, 
which we call solitons and antisolitons, respectively, 
defined by 
\begin{eqnarray}
 \mu_k =  \sqrt{\rho^2 + (z-a_k)^2} -(z-a_k) , \qquad
\bar \mu_k = - \sqrt{\rho^2 + (z-a_k)^2} -(z-a_k) \label{soliton},
\end{eqnarray}
where $a_k$ is a real constant.
We choose ${\mu'}_{\!\!k}$ $(i=1,\cdots,n)$ from either $\mu_k$ or $\bar\mu_k$.
We introduce $n$ 3-vectors $m^{(k)}$ 
associated with ${\mu'}_{\!\!k}$.
$m^{(k)}$ is called BZ vector.
We make a $n\times n$ matrix as
\begin{eqnarray}
&&\Gamma_{kl} =\frac{ m^{(k)}_a (\Psi_{g_0}^{-1} ({\mu'}_{\!\!k},\rho,z) )^{ab}
 g_{0bc} (\Psi_{g_0}^{-1} ({\mu'}_{\!\!l},\rho,z) )^{cd} m^{(l)}_d}
 {\rho^2 +{\mu'}_{\!\!k} {\mu'}_{\!\!l}}.\label{Gamma}
\end{eqnarray}
Then a metric
\begin{eqnarray}
&& {g'}_{\!\!1ab}= 
\left( g_{0ab} - \sum_{kl} (\Gamma^{-1})_{kl}{{\mu'}_{\!\!k}}^{-1}{{\mu'}_{\!\!l}}^{-1}
N^{(k)}_a N^{(l)}_b
\right) ,  \label{hatg}
\\
&&N^{(k)}_a\equiv m^{(k)}_b (\Psi_0^{-1} ({\mu'}_{\!\!k},\rho,z) )^{bc} g_{0ca}.\label{N}
\end{eqnarray}
satisfies the Einstein equations.
In general, ${\bf{ g'}_{\!\!1}}$ does not satisfies $\det {\bf g}=-\rho^2$.
However, ${\bf {g'}_{\!\!1}}$ multiplied by $\rho$ and ${\mu'}_{\!\!i}$
also satisfies the Einstein equations.
With the help of this property, 
we can get the metric satisfying both the Einstein equations
and $\det {\bf g_1} =-\rho^2$ as
\begin{eqnarray}
g_{1ab}= \rho^{-\frac{2n}{D}} \prod _{k=1}^n {{\mu'}_{\!\!k}}^{\frac{2}{D}}  {g'}_{\!\!1ab}.
 \label{newg}
\end{eqnarray}

Moreover, if we choose $f$ as
\begin{eqnarray}
f=C f_0 \rho^{\frac{-(n^2 +2n -Dn)}{D}} 
\left(\prod_{k=1}^n {{\mu'}_{\!\!k}}^{\frac{2(n+D-1)}{D}}\right)
\left(\prod_{k=1}^n ({{\mu'}_{\!\!k}}^2 \rho^2)^{\frac{2-D}{D}}\right)
\left(\prod _{k,l=1,k>l}^n ({\mu'}_{\!\!k} -{\mu'}_{\!\!l})^{\frac{4}{D}}\right)^{-1} 
\det \Gamma_{kl}, \label{newf}
\end{eqnarray}
with a constant $C$, 
$f$  satisfies eq.(\ref{rhof}) and eq.(\ref{zf}).
The above set of operations is called soliton transformation with \{${\mu'}_{\!\!i}$\}.

In this paper  
we use the strategy taken in Ref.\cite{Pomeransky2}.
Using this method, 
a new metric ${\bf g}$ automatically satisfies $\det {\bf g}=-\rho^2$ 
without the operation of eq.(\ref{newg}).
We give a diagonal seed metric ${\bf g_0}$ and $f_0$ which is constructed using $\mu_k$ and $\rho$ 
as explained in the next section.
First, we apply the soliton transformation 
with \{${\mu'}_{\!\!i}$\}
and simple BZ vectors to the seed metric ${\bf g_0}$, 
to obtain the metric $ {\bf g'_1}$ 
(see eq.(\ref{hatg})).
If we operate the soliton transformation to the metric ${\bf g'_1}$ 
with \{${{\bar\mu}'}_i$\}
and the same BZ vectors, 
the metric is transformed back into the seed metric ${\bf g_0}$ 
where we refer to $\bar\mu_i$ ($\mu_i$) as ${{\bar\mu}'}_{i}$ 
when ${\mu'}_{\!\!i}$ is $\mu_i$ ($\bar\mu_i$). 
If we operate the soliton transformation to the metric ${\bf g'_0}$ 
with \{${{\bar\mu}'}_i$\}
and general BZ vectors, 
we get a new metric ${\bf g}$.
Since the change of $\det {\bf g}$ is independent of BZ vectors,
change of $\det {\bf g}$ by soliton transformation from ${\bf g'_1}$ to ${\bf g_0}$
is the same as that from ${\bf g'_1}$ to ${\bf g}$.
This means $\det {\bf g}=\det {\bf g_0}$ and 
we don't need to rescale the new metric.
Moreover, we don't care the complicated factors composed of $\rho$ and $\mu'_{k}$ in eq.(\ref{newf}).
As a result, the metric ${\bf g}$ and $f$
take a simple form as
\begin{eqnarray}
&&\!\!\!\!\!\!
g_{ab}=  {g'}_{\!\!1ab} - \sum_{kl} ({\Gamma'}^{-1})_{kl}{\mu'}_{\!\!k}^{-1}{\mu'}_{\!\!l}^{-1}
N^{(k)}_a N^{(l)}_b 
 ,\\
&&
f=C f_0 \frac{\det {\Gamma'}_{kl}}{\det {\Gamma'}^{(0)}_{kl}} ,\label{f}
\end{eqnarray}
where ${\Gamma'}_{kl}$ and ${\Gamma'}_{kl}^{(0)}$ are constructed 
from the metric $g'_1$, and $\Psi_{g'_1}$ 
with general BZ vectors
the simple ones, respectively.

In the inverse scattering method 
regular points where $\det {\bf g} \neq 0$ are transformed to regular points. 
Some points where $\det {\bf g} =0$ become physical singularities.
However, we don't need to care about  physical singularities at the level of the seed metric
because physical singularities can be transformed to  coordinate singularities.  
In many cases, the seed metric from which a new metric 
without a physical singularity is obtained has 
physical singularities on the axis.

\section{Seed metric}\label{seed}

In this section,
we will give a seed solution 
which is one of the Weyl solutions~\cite{Emparan2}.
We have introduced solitons $\mu_k$ in eq.(\ref{soliton}).
Using these solitons, we can construct a solution of the Einstein equations as follows.
We prepare the $3\times3$ diagonal metric like
\begin{eqnarray}
{\bf g}=\mbox{diag} \left\{ 
-\frac{\mu_{i_1^1} \mu_{j^1_2} \cdots }{\mu_{{i'}^1_1}\cdots} ,
\frac{\rho^2 \mu_{i^2_1} \cdots}{ \mu_{{i'}^2_1} \cdots},
\frac{\mu_{i^3_1} \cdots}{\mu_{{i'}^3_1} \cdots}
\right\}, \label{gg}
\end{eqnarray}
where the total number of each $\mu_k$ in the numerator 
of all components is equal to that in the denominator.
Though in the above example $\rho^2$ appears in the second component,
it can be in any component but it appears only once.
You can check that the metric ${\bf g}$ satisfies Einstein equations (\ref{diffg}) and 
$\det{\bf g} = -\rho^2$ (See Appendix \ref{apg}).
Next, we give $f$ as
\begin{eqnarray}
&&f=\kappa^2 \left( \prod_{k} \mu_{i^2_k}\right)
\left(\prod_{k} \mu_{{i'}^2_k}^{-1} \right)
\prod_{a=1}^3\Biggl\{
\left(\prod_{k,l}\frac{ \rho^2+\mu_{i^a_k}\mu_{{i'}^a_l}}{\mu_{i^a_k}\mu_{{i'}^a_l}}\right)
\left(\prod_{k\neq l}\frac{ \rho^2+\mu_{i^a_k}\mu_{i^a_l}}{\mu_{i^a_k}\mu_{i^a_l}}\right)^{-1}
\left(\prod_{k\neq l}\frac{ \rho^2+\mu_{{i'}^a_k}\mu_{{i'}^a_l}}{\mu_{{i'}^a_k}\mu_{{j'}^a_l}}\right)^{-1}
\nonumber\\
&&\qquad\qquad\quad\qquad\qquad\qquad\qquad\qquad\qquad\qquad\qquad\qquad\times
\left(\prod_{k}\frac{ \rho^2+\mu_{i^a_k}^2}{\mu_{i^a_k}^2}\right)^{(-1/2)}
\left(\prod_{k}\frac{ \rho^2+\mu_{{i'}^a_k}^2}{\mu_{{i'}^a_k}^2}\right)^{(-1/2)}
\Biggr\}, \label{F}
\end{eqnarray}
where $\kappa$ is a constant 
and $\mu_{i^a_k}$ ($\mu_{{i'}^a_k}$) are 
the solitons that appear in the numerator (denominator) of the $a$-th component. 
Although, in the above example, $f$ has the product 
$\prod_{k} \mu_{i^2_k}\prod_{k} \mu_{{i'}^2_k}^{-1}$, 
we must replace it with the product of $\mu_{i^a_k}$
corresponding to the component having $\rho^2$ in the general case.
Then, $f$ satisfies Einstein equations (\ref{rhof}) and (\ref{zf}) (See Appendix \ref{apg}).
Since $\mu_i>0$ except for the rod,
the seed metric is regular in the region satisfying $\rho>0$.
Although this seed metric generally has singularities on the rod, 
it is not a problem at all  
as we explained above.

Using this prescription, we will show a method
for constructing a seed metric corresponding to a given rod structure.
Before that, we give a brief explanation about a rod structure. 
Since $\det {\bf g} =-\rho^2$,
at least one eigenvalue of ${\bf g}$ becomes zero at $\rho = 0$.
An eigenvector with a zero eigenvalue is not always the same but depend on $z$.
A rod structure represents how the eigenvector changes depending on $z$.
The positive density rod indicates the direction of the eigenvector 
with a zero eigenvalue.
In this paper, we consider only the case in which
the eigenvalues corresponding to positive density rods 
become $O(\rho^2)$ in the limit $\rho\to 0$.
Then, if there are two positive density rods, 
one of the other eigenvalue of ${\bf g}$ must become infinite 
as $O(\rho^{-2})$ in the limit $\rho \to0$.
We call it a negative density rod.

A method for constructing a seed metric corresponding to a given rod structure
is given as follows.
First, we put a minus sign to $t$-$t$ component and
$\rho^2$ to the numerator of
the component having rod at the left-end ($z=-\infty$).
Next, starting with the left-end,
we add $\mu_i$ to the numerator of the corresponding diagonal component
if the positive density rod appears at $a_i$ or 
if the negative density rod disappears at $a_i$.
Similarly, we add $\mu_i$ to the denominator of 
the corresponding diagonal component
if the positive density rod disappears at $a_i$ or 
if the negative density rod appears at $a_i$.
Constructing $f$ by eq.(\ref{F}),
we complete the construction of a seed metric. 
The metric ${\bf g}$ obtained by the above operations
has to the rod structure.
Suppose a positive (or negative) rod exists on the left of $a_i$.
When the value of $z$ crosses $a_i$ from the left,
the leading power in $\rho$ of $\mu_i$ changes from $O(1)$ to $O(\rho^2)$ near the rod.
Since the component corresponding to the positive (negative) density rod 
should behave as $O(\rho^2)$ ($O(\rho^{-2})$), 
we can make the positive (negative) density rod 
appearing from $z=a_i$
by adding $\mu_i$ to the numerator (denominator)
of the metric component.  
The disappearance of rod can be explained similarly.

\begin{figure}[tbp]
  \begin{center}
    \includegraphics[keepaspectratio=true,height=25mm]{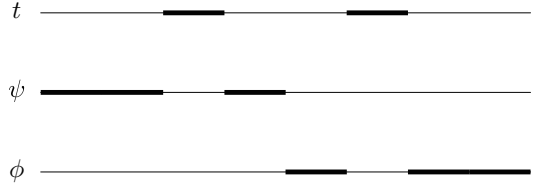}
  \end{center}
  \caption{the rod structure of the orthogonal black di-ring solution we want.}
  \label{fig:BR4.eps}
\end{figure}

\begin{figure}[tbp]
  \begin{center}
    \includegraphics[keepaspectratio=true,height=30mm]{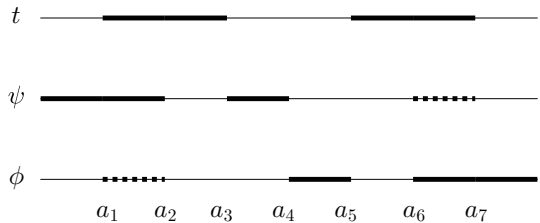}
  \end{center}
  \caption{the rod structure of the seed metric of the orthogonal black di-ring
             the solid (dashed) lines means positive (negative) density rods.}
  \label{fig:BR1.eps}
\end{figure}

Next, we construct the seed metric of the orthogonal black di-ring.
The rod structure of the orthogonal black di-ring solution is as shown in fig.\ref{fig:BR4.eps}.
The thick solid lines correspond to the positive density rods,
while the dashed lines correspond to the negative density rods.
Naively one may think that we need to prepare the seed metric as the one corresponding 
to this rod structure directly.
However, referring to the construction 
of the black saturn solution~\cite{saturn},
we expect that we cannot add the angular momentum of $S^1$-plane of black ring 
starting from the seed metric of this rod structure.
Following the case of black saturn,
we introduce a negative density rod as shown in fig.\ref{fig:BR1.eps}.
Then the metric corresponding to this rod structure can be written as
\begin{eqnarray}
{\bf g_0}=\mbox{diag} \left\{ 
-\frac{\mu_1 \mu_5}{\mu_3\mu_7} ,
\frac{\rho^2 \mu_3 \mu_7}{ \mu_2 \mu_4 \mu_6},
\frac{\mu_2\mu_4\mu_6}{\mu_1\mu_5}
\right\},\label{seedg}
\end{eqnarray}
where the constants $a_i$ contained in $\mu_i$ are ordered such that
 $a_i \ge  a_j$ for $i>j$.
Then $f_0$ becomes 
\begin{eqnarray}
&&f_0=k^2 \frac{\mu_2\mu_4\mu_6}{\mu_1\mu_5} (\rho^2 +\mu_1\mu_2)(\rho^2 +\mu_1\mu_3)
(\rho^2 +\mu_1\mu_4)
(\rho^2 +\mu_1\mu_6)(\rho^2 +\mu_1\mu_7)(\rho^2 +\mu_2\mu_3)
(\rho^2 +\mu_2\mu_5)
\nonumber\\
&&\qquad\times
(\rho^2 +\mu_2\mu_7)(\rho^2 +\mu_3\mu_4)(\rho^2 +\mu_3\mu_5)
(\rho^2 +\mu_3\mu_6)
(\rho^2 +\mu_4\mu_5)(\rho^2 +\mu_4\mu_7)(\rho^2 +\mu_5\mu_6)
(\rho^2 +\mu_5\mu_7)
\nonumber\\
&&\qquad\times
(\rho^2 +\mu_6\mu_7)
(\rho^2 +\mu_1\mu_5)^{-2}(\rho^2 +\mu_2\mu_4)^{-2}
(\rho^2 +\mu_2\mu_6)^{-2}
(\rho^2 +\mu_3\mu_7)^{-2}(\rho^2 +\mu_4\mu_6)^{-2}
\prod_{i=1}^7 (\rho^2 +\mu_i^2)^{-1} 
.\label{seedf}
\end{eqnarray}
This solution has singularities on the rods $z\in[a_1,a_2]$ and $z\in[a_6,a_7]$ ,and 
is of no physical interest by itself.
However, once soliton transformation is applied to this solution appropriately,
the negative density rod moves to the $t$-direction
and cancels the positive density rod
on the segments $z \in [a_1 ,a_2]$ and $z\in [a_6 , a_7]$.
Although in general it leaves the singularities at $z=a_1$ and $z=a_7$,
we will find that we can remove these singularities 
by choosing the appropriate BZ vectors. 

We must construct the matrix $\Psi_{g_0}$ 
solving eq.(\ref{psi1}) and eq. (\ref{psi2}).
Although it seems  at first sight difficult 
to solve these equations,
it is easy to obtain one of the solutions in fact.
The method is as follows. 
We eliminate $\rho^2$ using $\mu_i \bar \mu_i=-\rho^2$.
Here we can use any $\mu_i$ other than those used in soliton transformation\footnote{
You might worry that, if we replace $-\rho^2$ with $\mu_i \bar \mu_i$ of the different $\mu_i$, 
the different $\Psi$ is obtained. 
Then the difference of the component in $\Psi^{-1}(\lambda=\mu_k)$ is any factor.
In the new metric, $\Psi$ appears in the only form of $m_a^{(k)}\Psi^{-1}_{ab}(\mu_k)$. 
These facts mean the difference of $\Psi$ can be finally absorbed in the parameters in BZ vectors.}.
Then basically we only have to change $ \mu_i$ in ${\bf g_0}$ into 
$( \mu_i- \lambda)$ to get $\Psi_{g_0}$. 
However, if we perform a soliton transformation with $ \mu_i$ when ${\bf g_0}$ has $ \mu_i$, 
$\Psi_{g_0}( \mu_i,\rho,z)$ or $\Psi_{g_0}^{-1}( \mu_i,\rho,z)$ 
becomes infinity.
In order to avoid this pathology, 
we replace $\mu_i$ contained in ${\bf g_0}$  
with $-\mu_k \bar\mu_k / \bar\mu_i$ .
Then, 
we obtain the matrix $\Psi_{g_0}(\lambda,\rho,z)$ which has no pathology 
at $ \lambda= \mu_i$. 

\section{soliton transformation}\label{trans}

In this section, we explain the soliton transformation 
to obtain the orthogonal black di-ring solution 
following the construction of the black saturn solution~\cite{saturn}.
The orthogonal black di-ring solution can be obtained by 
performing the following soliton transformation on the
seed metric (\ref{seedg}).

We perform a 2-soliton transformation\footnote{
The metric obtained by 2-soliton transformation is the same as that 
obtained by two 1-soliton transformations with the same solitons 
and the same BZ vectors. 
The calculation of two 1-soliton transformations is easier than that of 
2-soliton transformation.} 
with
$\mu'_{i}=\{\mu_1 ,\bar\mu_7\}$ and both of BZ vectors being
$(1,0,0)$.
The metric obtained by this transformation is
\begin{eqnarray}
{\bf g'_{1}}= \mbox{diag} \left\{ 
-\frac{\mu_5 \mu_7}{\mu_3\mu_1} ,
\frac{\rho^2 \mu_3 \mu_7}{ \mu_2 \mu_4 \mu_6},
\frac{\mu_2\mu_4\mu_6}{\mu_1\mu_5}
\right\}.
\end{eqnarray}
We rescale this metric as
\begin{eqnarray}
{\bf {\tilde g'}_{1}} = \frac{\mu_1}{\mu_7} {\bf g'_{1}}=
 \mbox{diag} \left\{ 
-\frac{\mu_5}{\mu_3} ,
\frac{\rho^2 \mu_1 \mu_3}{ \mu_2 \mu_4 \mu_6},
\frac{\mu_2\mu_4\mu_6}{\mu_5\mu_7}
\right\}.\label{tildeg}
\end{eqnarray}
This rescaling makes the calculation easier,  
although it does not change the result\footnote{
Due to the rescaling, 
$\Psi$ is also rescaled as 
$\Psi'_0= (\mu_1-\lambda)/(\mu_7-\lambda)\tilde\Psi_0$.
This affect $\Gamma_{kl}$ of eq.(\ref{Gamma}) and $N^{(k)}$ of eq.(\ref{N}) as 
$\Gamma_{kl} \to (\mu_1-\mu_k)(\mu_1-\mu_l)/((\mu_7-\mu_k)(\mu_7-\mu_l))\Gamma_{kl}$ 
and $N^{(k)} \to (\mu_1-\mu_k)/(\mu_7-\mu_k) N^{(k)}$.
We can find from eq.(\ref{hatg}) that in the new metric the effects cancel each other. 
}. 
We make $\Psi$ corresponding to ${\bf \tilde g'_1}$ as
\begin{eqnarray}
&&
\Psi_{\tilde g'_1} =
 \mbox{diag}\biggl\{
-\frac{(\mu_5-\lambda)}{(\mu_3-\lambda)} ,
-\frac{ (\mu_1-\lambda) (\mu_3-\lambda)(\bar\mu_4-\lambda)}
{ (\mu_2-\lambda) (\mu_6-\lambda)}
,\frac{(\mu_2-\lambda)(\mu_6-\lambda)(\bar\mu_7-\lambda)}
{(\mu_5-\lambda)(\bar\mu_4-\lambda)}
\biggr\}.
\end{eqnarray}
Next, we apply the 2-soliton transformation to ${\bf \tilde g'}$ with $\bar \mu'_i$.
The BZ vectors associated with $\bar\mu_1$ and $\mu_7$ are chosen to be
(1,0,c) and
(1,b,0), respectively.
We denote the resulting metric by ${\bf \tilde g}$.
Finally, to undo the rescaling in eq.(\ref{tildeg})
we rescale ${\bf \tilde g}$ as
\begin{eqnarray}
{\bf g}= \frac{\mu_7}{\mu_1}{\bf \tilde g}.
\end{eqnarray}

We must also compute $f$ given in eq.(\ref{f}).
$\Gamma_{kl}$, which is necessary to compute eq.(\ref{f}), 
is obtained in the process of constructing ${\bf \tilde g}$.
Moreover, $\Gamma_{kl}^{(0)}$ is given by 
$\Gamma_{kl}^{(0)} =\Gamma_{kl}|_{b=0,c=0}$.


The obtained metric 
is given by eq.(\ref{metric}) with
$a,b=t,\psi,\phi$ and

\begin{eqnarray}
&&
f=Cf_0 H/F,
\\
&&
H=F +b^2 F_{(b)} +c^2 F_{(c)} + b^2 c^2 F_{{(bc)}}, 
\\
&&
F = -\frac{\mu_5^2(\mu_3-\mu_7)^2(\mu_1 \mu_3 +\rho^2)^2
(\mu_1 \mu_7 +\rho^2)^2}{\mu_3^2 (\mu_5-\mu_7)^2
(\mu_1\mu_5 +\rho^2)^2 (\mu_1^2 +\rho^2)(\mu_7^2+\rho^2)\rho^4},
\\
&&
F_{(b)}=-\frac{\mu_1\mu_4\mu_5(\mu_2-\mu_7)^2
(\mu_6-\mu_7)^2(\mu_1\mu_3+\rho^2)^2}
{\mu_2\mu_6(\mu_3-\mu_7)^2
(\mu_1\mu_5 +\rho^2)^2(\mu_4\mu_7+\rho^2)^2
(\mu_1^2+\rho^2)(\mu_7^2+\rho^2)},
\\
&&
F_{(c)}=-\frac{\mu_1^2 \mu_2 \mu_6 \mu_7 
(\mu_1-\mu_4)^2(\mu_3-\mu_7)^2 (\mu_1\mu_5+\rho^2)^2}
{\mu_3\mu_4(\mu_5-\mu_7)^2
(\mu_1\mu_2 +\rho^2)^2(\mu_1\mu_6+\rho^2)^2
(\mu_1^2+\rho^2)(\mu_7^2+\rho^2)\rho^2},
\\
&&
F_{(bc)}=\frac{\mu_1^3 \mu_3 \mu_7 
(\mu_1-\mu_4)^2(\mu_2-\mu_7)^2
(\mu_6-\mu_7)^2 (\mu_1\mu_5+\rho^2)^2}
{\mu_5(\mu_1-\mu_7)^2(\mu_3-\mu_7)^2
(\mu_1\mu_2 +\rho^2)^2(\mu_1\mu_6+\rho^2)^2(\mu_4\mu_7+\rho^2)^2
(\mu_1^2+\rho^2)(\mu_7^2+\rho^2)},
\end{eqnarray}
\begin{eqnarray}
&&
g_{tt}=H^{-1}(A +b^2 A_{(b)} +c^2 A_{(c)} +b^2 c^2 A_{(bc)}),
\\
&&
A=\frac{\mu_1 \mu_5^3(\mu_3-\mu_7)^2(\mu_1\mu_3+\rho^2)^2
(\mu_1\mu_7+\rho^2)^2}
{\mu_3^3\mu_7(\mu_5-\mu_7)^2
(\mu_1\mu_5+\rho^2)^2(\mu_1^2+\rho^2)(\mu_7^2+\rho^2)\rho^4},
\\
&&
A_{(b)}=-\frac{\mu_1^2 \mu_4 \mu_5^2\mu_7(\mu_2-\mu_7)^2(\mu_6-\mu_7)^2
(\mu_1\mu_3+\rho^2)^2}
{\mu_2\mu_3\mu_6(\mu_3-\mu_7)^2
(\mu_1\mu_5+\rho^2)^2(\mu_4\mu_7+\rho^2)^2
(\mu_1^2+\rho^2)(\mu_7^2+\rho^2)\rho^2},
\\
&&
A_{(c)}=-\frac{\mu_1 \mu_2 \mu_5 \mu_6(\mu_1-\mu_4)^2(\mu_3-\mu_7)^2
(\mu_1\mu_5+\rho^2)^2}
{\mu_3^2\mu_4(\mu_5-\mu_7)^2
(\mu_1\mu_2+\rho^2)^2(\mu_1\mu_6+\rho^2)^2
(\mu_1^2+\rho^2)(\mu_7^2+\rho^2)},
\\
&&
A_{(bc)}=-\frac{\mu_1^2 \mu_7^2(\mu_1-\mu_4)^2(\mu_2-\mu_7)^2
(\mu_6-\mu_7)^2(\mu_1\mu_5+\rho^2)^2}
{(\mu_1-\mu_7)^2(\mu_3-\mu_7)^2
(\mu_1\mu_2+\rho^2)^2(\mu_1\mu_6+\rho^2)^2(\mu_4\mu_7+\rho^2)^2
(\mu_1^2+\rho^2)(\mu_7^2+\rho^2)},
\end{eqnarray}
\begin{eqnarray}
&&
g_{\psi\psi}= B+H^{-1}(b^2 B_{(b)}  +b^2 c^2 B_{(bc)}),
\\
&&
B=\frac{\mu_3 \mu_7 \rho^2}
{\mu_2 \mu_4 \mu_6},
\\
&&
B_{(b)}=\frac{\mu_1 \mu_3\mu_5(\mu_2-\mu_7)^2(\mu_6-\mu_7)^2
(\mu_1\mu_3+\rho^2)^2 \rho^2}
{\mu_2^2\mu_6^2\mu_7(\mu_3-\mu_7)^2
(\mu_1\mu_5+\rho^2)^2(\mu_4\mu_7+\rho^2)^2(\mu_1^2+\rho^2)},
\\
&&
B_{(bc)}=-\frac{\mu_1^3 \mu_3^2 (\mu_1-\mu_4)^2(\mu_2-\mu_7)^2
(\mu_6-\mu_7)^2 (\mu_1\mu_5+\rho^2)^2 \rho^2}
{\mu_2\mu_4\mu_5\mu_6(\mu_1-\mu_7)^2(\mu_3-\mu_7)^2
(\mu_1\mu_2+\rho^2)^2(\mu_1\mu_6+\rho^2)^2
(\mu_4\mu_7+\rho^2)^2(\mu_1^2+\rho^2)},
\end{eqnarray}

\begin{eqnarray}
&&
g_{\phi\phi}=C+ H^{-1}(c^2 C_{(c)}  +b^2 c^2 C_{(bc)}),
\\
&&
C=\frac{\mu_2 \mu_4 \mu_6}
{\mu_1 \mu_5},
\\
&&
C_{(c)}=\frac{\mu_1 \mu_2^2\mu_6^2\mu_7(\mu_1-\mu_4)^2(\mu_3-\mu_7)^2
(\mu_1\mu_5+\rho^2)^2 }
{\mu_3\mu_5(\mu_5-\mu_7)^2(\mu_1\mu_2+\rho^2)^2
(\mu_1\mu_6+\rho^2)^2(\mu_7^2+\rho^2) \rho^4},
\\
&&
C_{(bc)}=-\frac{\mu_1^2 \mu_2\mu_3\mu_4\mu_6\mu_7(\mu_1-\mu_4)^2(\mu_2-\mu_7)^2
(\mu_6-\mu_7)^2(\mu_1\mu_5+\rho^2)^2 }
{\mu_5^2(\mu_1-\mu_7)^2(\mu_3-\mu_7)^2(\mu_1\mu_2+\rho^2)^2
(\mu_1\mu_6+\rho^2)^2(\mu_4\mu_7+\rho^2)^2(\mu_7^2+\rho^2) \rho^2},
\end{eqnarray}

\begin{eqnarray}
&&
g_{t\psi} = H^{-1} b(D +c^2 D_{(c)}),
\\
&&
D=\frac{\mu_1 \mu_5^2 (\mu_2-\mu_7)(\mu_6-\mu_7)
(\mu_1\mu_3+\rho^2)^2 (\mu_1\mu_7+\rho^2)}
{\mu_2\mu_3\mu_6\mu_7(\mu_5-\mu_7)
(\mu_1\mu_5+\rho^2)^2(\mu_4\mu_7+\rho^2) (\mu_1^2+\rho^2) \rho^2},
\\
&&
D_{(c)}=-\frac{\mu_1^2 (\mu_1-\mu_4)^2 (\mu_2-\mu_7)(\mu_6-\mu_7)
(\mu_1\mu_5+\rho^2)^2 }
{\mu_4(\mu_1-\mu_7)(\mu_5-\mu_7)
(\mu_1\mu_2+\rho^2)^2(\mu_1\mu_6+\rho^2)^2
(\mu_4\mu_7+\rho^2)(\mu_1^2+\rho^2)},
\end{eqnarray}

\begin{eqnarray}
&&
g_{t\phi} = H^{-1} c(E +b^2 E_{(b)}),
\\
&&
E=\frac{ \mu_2\mu_5\mu_6 (\mu_1-\mu_4)(\mu_3-\mu_7)^2
(\mu_1\mu_3+\rho^2) (\mu_1\mu_7+\rho^2)}
{\mu_3^2(\mu_5-\mu_7)^2(\mu_1\mu_2+\rho^2)
(\mu_1\mu_6+\rho^2)(\mu_7^2+ \rho^2)\rho^4},
\\
&&
E_{(b)}=-\frac{\mu_1 \mu_4\mu_7 (\mu_1-\mu_4) (\mu_2-\mu_7)^2
(\mu_6-\mu_7)^2 (\mu_1\mu_3+\rho^2) }
{(\mu_1-\mu_7)(\mu_3-\mu_7)^2(\mu_1\mu_2+\rho^2)
(\mu_1\mu_6+\rho^2)(\mu_4\mu_7+\rho^2)^2(\mu_7^2+\rho^2)\rho^2 },
\end{eqnarray}

\begin{eqnarray}
g_{\psi\phi}=H^{-1} bc \frac{\mu_1 (\mu_1-\mu_4) (\mu_2-\mu_7)
(\mu_6-\mu_7) (\mu_1\mu_3+\rho^2) }
{(\mu_1-\mu_7)(\mu_5-\mu_7)(\mu_1\mu_2+\rho^2)
(\mu_1\mu_6+\rho^2)(\mu_4\mu_7+\rho^2)\rho^2 }.
\end{eqnarray}

This solution has generally point singularities at 
$(\rho,z)=(0,a_1)$ and $(0,a_7)$
and conical singularities on the rods $z \in[-\infty,a_2]$,
$z \in[a_3,a_4]$, $z \in[a_4,a_5]$ and $z \in[a_6,\infty]$.
However, in the next section, we show
these singularities can be removed if we choose the parameters appropriately.
When we remove the singularities, 
this solution becomes the orthonormal black di-ring solution.

\section{analysis}\label{analysis}

First, we analyse the rod structure of the orthogonal black di-ring solution.
Although  
generally singularities appear at $z=a_1$ and $a_7$,
we can remove these singularities by choosing parameters $c$ and $b$
appropriately. 
After setting the parameters $b$ and $c$ to the particular values that 
eliminate the singularities at $z=a_1$ and $a_7$,  
we next analyse the asymptotic behavior of this solution.
If we choose $C \kappa^2=1$, we will find that this solution becomes asymptotically flat
Finally, we analyse the conical structure around the axis.
We will find that all conical singularities can be removed by choosing 
the parameters appropriately.
As a result, four parameters are left in the end.

\subsection{rod structure}

\begin{figure}[tbp]
  \begin{center}
    \includegraphics[keepaspectratio=true,height=30mm]{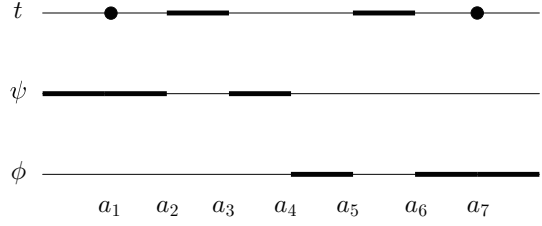}
  \end{center}
  \caption{General rod structure after the transformation.
          Although this solution has the singularities at $z=a_1$ and $z=a_7$, 
          we can remove these singularities 
          if we choose the appropriate parameters.}
  \label{fig:BR2.eps}
\end{figure}

The rod structure of the obtained metric with general parameters 
is illustrated in fig.\ref{fig:BR2.eps}.
We calculate the direction of the rod 
which is defined in Ref.\cite{Harmark}.
The direction of the rod represents the vector whose norm 
becomes zero at $\rho=0$. 
The semi-infinite rod $z\in [-\infty,a_2]$ and 
the finite rod $z[a_3,a_4]$ have directions $(0,1,0)$ 
which correspond to the $\psi$-axis. 
The semi-infinite rod $z\in [a_6,\infty]$ and 
the finite rod $z[a_4,a_5]$ have directions $(0,0,1)$ 
which correspond to the $\phi$-axis.
The finite rods $z[a_2,a_3]$ and $z[a_5,a_6]$, which 
correspond to the location of the black ring horizons,  
have directions
\begin{eqnarray}
v=(1,\Omega_\Psi^{(1)},\Omega_\Phi^{(1)}) \qquad \mbox{and}
\qquad u=(1,\Omega_\Psi^{(2)},\Omega_\Phi^{(2)}), \label{v}
\end{eqnarray}
respectively,
where 
\begin{eqnarray}
&&\Omega_\psi^{(1)}=-\frac{b\, a_{75}\, a_{76}}{2\, a_{73}^{\ 2}\, a_{71}},\qquad
\Omega_\phi^{(1)}=-\frac{c\, a_{51}^{\ 2}}{2\, a_{31}\, a_{61}\, a_{71}},
\\
&&\Omega_\psi^{(2)}=-\frac{b\, a_{76}}{2\, a_{71}\, a_{74}}, \qquad
\Omega_\phi^{(2)}=-\frac{c\, a_{41}}{2\, a_{61}\, a_{71}},
\\
&&
a_{ij}\equiv a_i-a_i.
\end{eqnarray}
This solution has no negative density rod as anticipated.
The negative density rod in the seed metric was cancelled by the positive density rod in $t$-direction.
However, the soliton transformation leaves the singularities at
$z=a_1$ and $a_7$. 
We find that the metric component $g_{\phi\phi}$ becomes the singular at $z=a_1$.
In order to see this, 
we perform the  coordinate transformation, 
\begin{eqnarray}
\rho= r \sin \theta, \qquad z = r \cos \theta + a_1.
\end{eqnarray}
In the limit $r \to 0$ (which means $\rho \to 0$ and 
$z\to a_1$), the leading term of $g_{\phi\phi}$ becomes 
\begin{eqnarray}
g_{\phi\phi} \to 4
\frac{a_{21}\,  a_{41}\,  a_{61}\, (-c^2\,  a_{41}\, a_{51}^{\ 2}\,
  +2\,  a_{21}\,  a_{31}\,  a_{61}\,  a_{71})\sin\theta(1-\cos\theta)}
{c^2\,  a_{41}\,  a_{51}^{\ 3}\,  + 2\, a_{12}\,  a_{13}\, 
 a_{51}\,  a_{61}\,  a_{71} \sin^2\theta (1-\cos\theta)^2} r^{-1}.
\end{eqnarray}
Obviously, $g_{\phi\phi}\to\infty$ for $r\to 0$.
However, this singularity can be removed 
if we choose the parameter as 
\begin{eqnarray}
c^2 = \frac{2\, a_{21}\, a_{31}\, a_{61}\, a_{71}}
{a_{41}\, a_{51}^{\ 2}}. \label{c}
\end{eqnarray}
Then the term proportion to $r^{-1}$ in $g_{\phi\phi}$ becomes $0$, 
and the leading term of $g_{\phi\phi}$ becomes
\begin{eqnarray}
g_{\phi\phi} \to 
2\frac{a_{31}\, a_{41}\, a_{51}\, a_{61}\, a_{71}
+a_{21}\, a_{41}\, a_{61}\, a_{71}\, a_{53}
+a_{21}\, a_{31}\, a_{61}\, a_{71}\, a_{54}
+a_{21}\, a_{31}\, a_{41}\, a_{51}\, a_{61}
+a_{21}\, a_{31}\, a_{41}\, a_{51}\, a_{71}}
{a_{31}\, a_{51}^{\ 2}\, a_{71}}.
\end{eqnarray}
In this limit, the other components of $g$ and $f$ become
\begin{eqnarray}
&&g_{tt}=O(r), \qquad
g_{\psi\psi}\to \frac{a_{31}\, a_{71}}{2\, a_{21}\, a_{41}\, a_{61}} r^2 \sin^2\theta,
\qquad 
g_{t\psi}=-\frac{b\,  a_{31}\, a_{72}\, a_{75}\, a_{76}}
{4\,  a_{21}\, a_{41}\, a_{61}\, a_{71}\, a_{73}^{\ 2}}
r^2 \sin^2 \theta,
\nonumber\\
&&
g_{t\phi}\to c\frac{a_{41}\, a_{51}}{a_{31}\, a_{71}}, 
\qquad
g_{\psi\phi}=\frac{c\,  a_{41}\, a_{51}}{a_{31}\, a_{71}} r^2 \sin^2 \theta \qquad 
\mbox{and} \qquad f\to \frac{a_{31}\, a_{71}}{2\, a_{21}\, a_{41}\, a_{61}}.
\end{eqnarray}
We find that, in this limit,
$g_{t\phi}$ and $g_{\phi\phi}$ are constant.
Moreover, when we introduce new variable 
\begin{eqnarray}
\psi_1= \psi-\frac{b\, a_{72}\, a_{75}\, a_{76}}
{2\, a_{71}^{\ 2}\, a_{73}^{\ 2}}t
+\frac{2\, c\, a_{21}\, a_{41}^{\ 2}\, a_{51}\, a_{61}}
{a_{31}^{\ 2}\, a_{71}^{\ 2}}\phi.,
\end{eqnarray}
the metric becomes
\begin{eqnarray}
ds^2 \simeq f(dr^2 +r^2 d\theta^2 + r^2 \sin^2\theta d{\psi_1}^2)
+2 g_{t\phi} dt d\phi +g_{\phi\phi} d\phi^2,
\end{eqnarray}
Since this metric is the Minkowski metric,
we find that there is no singularity at this point.

\begin{figure}[tbp]
  \begin{center}
    \includegraphics[keepaspectratio=true,height=33mm]{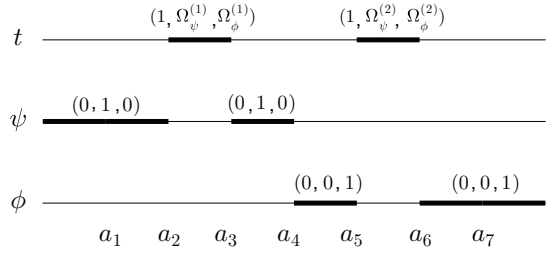}
  \end{center}
  \caption{The rod structure of the orthogonal black di-ring,
         which is obtained by removing the singularities by fixing $c$ and $b$.
         The vectors accompanied with rods are such defined in Ref.~\cite{Harmark}.}
  \label{fig:BR3.eps}
\end{figure}

In a similar way, we can remove the singularity at $z=a_7$
if we choose the parameter as\footnote{
If the seed metric is exchanged using $\mu_i\bar\mu_i=-\rho^2$, 
the result by a soliton transformation does not change
as we show the footnote-1.
When we write the seed metric with only $\bar\mu_i$ using $\mu_i\bar\mu_i=-\rho^2$,
the original seed metric and the transformed seed metric is symmetric as 
$\mu_i \to \bar \mu_{7-i}$ and $\psi \leftrightarrow \phi$.
From this, we find that, when we can remove a singularity at $z=a_1$,
we can have singularity appear at $z=a_7$.}
\begin{eqnarray}
&&b^2 = \frac{2\, a_{71}\, a_{73}^{\ 2}\, a_{74}}
{a_{72}\, a_{75}\, a_{76}}.\label{b}
\end{eqnarray}
With these conditions (\ref{c}) and (\ref{b}), 
the metric is smooth across $z=a_1$ and $a_7$.
Then the rod structure is as illustrated in fig.\ref{fig:BR3.eps}.

\subsection{asymptotic structure}

To analyse the asymptotic structure, 
we take other coordinates defined by
\begin{eqnarray}
\rho= \frac{1}{2} r^2 \sin{2\theta},
\qquad
z= \frac{1}{2} r^2 \cos{2\theta},
\end{eqnarray}
where $0< \theta< \pi/2$.
Then in the asymptotic limit $r \to \infty$, we have
$g_{t\psi}\to 0$, $g_{t\phi}\to 0$ and $g_{\psi\phi}\to 0$.
Since 
\begin{eqnarray}
d\rho^2 +dz^2 =r^2 (dr^2 +r^2 d\theta^2),
\end{eqnarray}
and the asymptotically value of $f$ becomes
\begin{eqnarray}
f=C\kappa^2\frac{1}{r^2} +\cdots,
\end{eqnarray}
we find we must choose $C\kappa^2=1$ 
in order to get the asymptotic flat solution .
Setting $C\kappa^2=1$ the asymptotic metric becomes
\begin{eqnarray}
ds^2 = -dt^2 +dr^2 + r^2 d\theta^2 
+r^2 \cos^2\theta d\psi^2
+r^2 \sin^2\theta d\phi^2.
\end{eqnarray}
The asymptotic flatness also requires 
that the angles $\psi$ and $\phi$ should have periodicities
\begin{eqnarray}
\Delta\psi =\Delta\phi =2\pi.
\end{eqnarray}

\subsection{regularities on the axis}

In order to avoid conical singularities,
the  period $\Delta\eta$ of the spacelike coordinate $\eta$($=\psi$ or $\phi$)
corresponding to the angle around each positive density rod
must satisfy~\cite{Harmark}
\begin{eqnarray}
\Delta\eta= 2\pi \lim_{\rho\to 0}
\sqrt{\frac{\rho^2 f}{g_{\eta\eta}}}.
\end{eqnarray}
When we impose the conditions (\ref{c}) and (\ref{b}),
the regularity conditions on the rods $z\in [-\infty,a_2]$
and  $z\in [a_6,\infty]$
are automatically satisfied for  $\Delta\psi=\Delta\phi=2\pi$

Next we consider the regularity condition on the rod 
$z\in [a_3,a_4]$.
The regularity condition is written as
\begin{eqnarray}
\Delta\psi = 2\pi \sqrt{\frac{
a_{14}\, a_{16}\, a_{17}\, a_{25}\, a_{27}\, a_{34}\, a_{35}\, a_{36}}
{a_{15}^{\ 2}\, a_{24}^{\ 2}\, a_{26}^{\ 2}\, a_{37}^{\ 2}}}.
\end{eqnarray}
Since $\Delta\psi$ must be $2\pi$,
this implies
\begin{eqnarray}
1 = \frac{
a_{14}\, a_{16}\, a_{17}\, a_{25}\, a_{27}\, a_{34}\, a_{35}\, a_{36}}
{a_{15}^{\ 2}\, a_{24}^{\ 2}\, a_{26}^{\ 2}\, a_{37}^{\ 2}}.
\label{regpsi}
\end{eqnarray}
Similarly, the regularity condition on the 
rod $z\in [a_4,a_5]$ 
implies 
\begin{eqnarray}
1 = \frac{a_{74}\, a_{72}\, a_{71}\, a_{63}\, a_{61}\, a_{54}\, a_{53}\, a_{52}}
{a_{73}^{\ 2}\, a_{64}^{\ 2}\, a_{62}^{\ 2}\, a_{51}^{\ 2}}.
\label{regphi}
\end{eqnarray}
Under the transformation, 
 $a_i \to a_{7-i}$,
eq.(\ref{regpsi}) becomes eq.(\ref{regphi}).
This is expected because 
the solution we want has a symmetry 
corresponding to the exchange of $\psi$ and $\phi$.
 
We must check the existence of the parameters 
which satisfy eq.(\ref{regpsi}) and  eq.(\ref{regphi})
with $a_i>a_j$ for $i>j$.
Without loss of generality, 
we can set $a_4 =0$.
For the sake of simplicity, 
we consider the symmetric case where 
\begin{eqnarray}
a_1 = -a_7, 
\qquad
a_2 =-a_6,
\qquad
a_3=-a_5.
\end{eqnarray}
In this case, eq.(\ref{regpsi}) and eq.(\ref{regphi})
become the same equation.
Moreover, we set $a_7 =1$, 
which corresponds to fixing the scale.
Then, eq.(\ref{regpsi}) and eq.(\ref{regphi}) 
are written as
\begin{eqnarray}
a_5^2(1+a_6)^2(a_5+a_6)^2 -(1+a_5)^4 a_6^4=0.
\label{reg}
\end{eqnarray}
Here, we denote the left hand side of this equation 
by $f(a_6)$ as a function of $a_6$.
Then
\begin{eqnarray}
&&f(a_5)=a_5^4 (1+a_5)^2(1-a_5)(3+a_5),
\label{a5}\\
&&f(1)= (1+a_5)^2(3a_5+1)(a_5-1).\label{1}
\end{eqnarray}
Since $a_5$ satisfies the condition 
$0<a_5<1$,
$f(a_5)$ is always positive.
On the other hand, 
$f(1)$ is always negative.
Therefore the parameter $a_6$ which satisfies eq.(\ref{reg}) exists 
between $a_5$ and $a_7(=1)$.
This means that the orthogonal black di-ring solution 
which has no singularity outside the horizon exists.

\subsection{regularities at end points of rods}

In this subsection, we show there is no singularity at the end points of rods.
From the symmetry of this solution (see footnote-3),
we only have to check the points $z=a_2$, $a_3$ and $a_4$.
Near each point $(\rho,z)=(0,a_i)$, we transform $\rho$ and $z$ as
\begin{eqnarray}
\rho=\frac{1}{2}r^2 \sin2\theta, \qquad z=\frac{1}{2}r^2 \cos2\theta -a_i.
\end{eqnarray}

Near the point $(\rho,z)=(0,a_2)$, 
we take basis as $v$ (see eq.(\ref{v})), $q=(0,1,0)$ and $s=(0,0,1)$.
In the limit $r \to 0$, inner products and $f$ become
\begin{eqnarray}
&&g(v,v)\to -4\frac{c^2\,  a_{42}\,  a_{62}\, a_{51}^{\ 4}}
{a_{31}\, a_{52}\, a_{61}\, a_{71}}r^2 \sin^2\theta,
\qquad
g(q,q) \to \frac{a_{32}\, a_{72}}{a_{42}\, a_{62}}r^2 \cos^2 \theta, \qquad
g(s,s) \to 4\frac{a_{21}\, a_{31}^{\ 2}\, a_{61}^{\ 2}\, a_{71}^{\ 2}\, a_{52}}
{c^2\,  a_{51}^{\ 4}\, a_{32}\, a_{72}},
\nonumber\\
&& \!\!\!\!
g(v,q)=O(r^4), \qquad
g(v,s)=\alpha_2 r^2 \sin^2\theta, \qquad
g(q,s)=\beta_2 r^2 \cos^2\theta, \qquad 
f\to \frac{c^2\,  a_{41}\, a_{51}^{\ 2}\, a_{32}\, a_{72}}
{2\, a_{21}\, a_{31}\, a_{61}\, a_{71}\, a_{42}\, a_{62}}r^{-2},
\end{eqnarray}
where $\alpha_2$ and $\beta_2$ are some constants.
We find that $g(s,s)$ is constant.
When bases are changed like 
\begin{eqnarray}
\psi_2=\psi+\frac{\beta_2}{f}\phi 
\qquad \mbox{and} \qquad 
\eta_2= \frac{1}{f}
\left(
4\frac{c^2 \, a_{42}\,  a_{62}\, a_{51}^{\ 4}}
{a_{31}\, a_{52}\, a_{61}\, a_{71}}\eta - \alpha_2\phi
\right) ,
\end{eqnarray}
where $\eta$ is defined as $(\partial/\partial \eta)^a\equiv v^a$ ,
the metric becomes
\begin{eqnarray}
ds^2 \simeq f(dr^2 +r^2 d\theta^2 + r^2 \cos^2\theta \psi_2^2-r^2 \sin^2\theta d\eta_2^2)
+ 4\frac{a_{21}\, a_{31}^{\ 2}\, a_{61}^{\ 2}\, a_{71}^{\ 2}\, a_{52}}
{c^2\,  a_{51}^{\ 4}\, a_{32}\, a_{72}} d\phi^2,\label{a2}
\end{eqnarray}
where we used eq.(\ref{c}).
Since eq.(\ref{a2}) is locally the Minkowski metric,
there is no singularity at the point $(\rho,z)=(0,a_2)$.

Similarly, we can obtain the metric near the point $(\rho,z)=(0,a_3)$.
The metric around this point becomes
\begin{eqnarray}
ds^2 \simeq \frac{a_{32}\, a_{73}}{a_{43}\, a_{63}}
(dr^2 +r^2 d\theta^2 + r^2 \sin^2\theta d\psi^2-r^2 \cos^2\theta d\eta_3^2)
+ 2\frac{a_{31}\, a_{43}\, a_{63}}{a_{32}\, a_{53}} d\phi^2,
\end{eqnarray}
where
\begin{eqnarray}
\eta_3= \frac{2\, a_{31}\, a_{43}\, a_{53}\, a_{63}}
{a_{32}\, a_{73}^{\ 2}}\eta.
\end{eqnarray}
From this, we find there is no singularity at the point $(\rho,z)=(0,a_3)$.

Finally we check the point at $(\rho,z)=(0,a_4)$.
The metric around this point becomes
\begin{eqnarray}
&&ds^2 \simeq \frac{a_{24}\, a_{74}}{a_{43}\, a_{63}}
(dr^2 +r^2 d\theta^2 + r^2 \sin^2\theta d\psi_4^2
+r^2 \cos^2\theta d\phi_4^2)
-\frac{a_{34}\, a_{54}}{a_{41}\, a_{74}} dt^2,
\nonumber\\
&&
\psi_4= \psi+ \frac{b\,  a_{41}\, a_{75}\, a_{76}}
{2a_{71}\, a_{64}\, a_{74}^{\ 2}}t
\qquad\mbox{and}\qquad
\phi_4=\phi+\frac{c\,  a_{51}^{\ 2}}{2\, a_{41}\, a_{61}\, a_{71}}t,
\end{eqnarray}
where we used eq.(\ref{regpsi}) and  eq.(\ref{regphi}).
We find that 
there is no singularity at the point $\rho=0$ and $z=a_4$.

\section{summary and discussion}\label{summary}

We present a method for constructing a seed  metric.
Using this method, we can obtain a seed metric
corresponding to any rod structure that we want.
In this paper, we constructed a solution which has two mutually
orthogonal black rings. 
We call this solution the orthogonal black di-ring.
Although the solution that we obtained by using the inverse scattering method 
generally has singularities,
we have shown that we can remove these singularities 
by choosing the parameters appropriately.
This solution has four free parameters 
which represent 
radii of black rings,
and speeds of $S^1$-rotation of black rings.

\begin{figure}[tbp]
  \begin{center}
    \includegraphics[keepaspectratio=true,height=25mm]{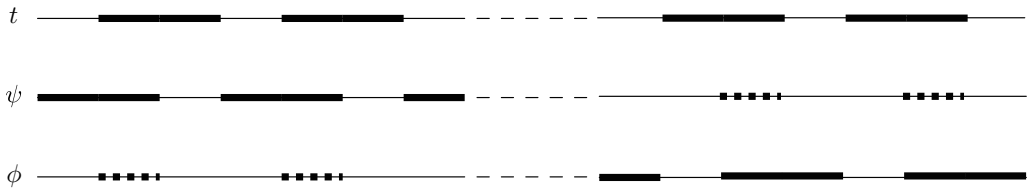}
  \end{center}
  \caption{The expected rod structure of the seed metric of the many 
  black rings solution.}
  \label{fig:BR5.eps}
\end{figure}

If we want to construct many black rings solution, 
probably we should prepare the seed metric corresponding to 
the rod structure as shown in fig.\ref{fig:BR5.eps},
which can be easily obtained by our method.
The soliton transformation applied to this seed metric
leaves naked singularities 
with the general parameters.
However, in generating the orthogonal black di-ring solution, 
the naked singularities can be removed 
when we choose the parameters by imposing the continuity of the periodicities at this point.
This seems to tell us that 
these singularities are related to the discontinuity of the periodicities.
The physical reason why the discontinuity of the periodicities exists in the seed metric is
that a plane which helps the balance against its attractive self-gravity force is needed.
If the rotation of the black ring balances against its attractive self-gravity force,
such a plane is not needed.
Since the rotation of the black ring is introduce through the BZ parameters, 
it seems that we can choose the parameters such that no discontinuity of the periodicities appears.
Therefore, the naked singularities can be probably removed.

Similarly, it seems that
the solution, which has black rings on the two orthogonal planes and a black hole at the center, 
can be probably obtained.
We will report such a solution in the forthcoming paper. 

\acknowledgements
The author is grateful to Takahiro Tanaka for careful reading the manuscript and 
useful comments.
He also thanks Takashi Nakamura for his valuable comments and continuous
encouragement.
This work is supported in part by the 21st Century COE ``Center for
Diversity
and Universality in Physics'' at Kyoto university, from the Ministry of
Education,
Culture, Sports, Science and Technology of Japan.\\

\appendix

\section{}\label{apg}

In this section, we show that eq.(\ref{gg}) and eq.(\ref{F})
satisfy the Einstein equations (\ref{diffg}-\ref{zf}) and $\det {\bf g}=-\rho^2$.
Thanks to the diagonalization, we can divide eq.(\ref{diffg}) 
into that for each component of ${\bf g}$.
The differential equation of $g_{tt}$ becomes 
\begin{eqnarray}
\partial_\rho \{\rho \partial_\rho (\ln g_{tt}) \}+
\partial_z \{\rho \partial_z (\ln g_{tt}) \}=0. \label{gtt}
\end{eqnarray}
Suppose that $g_{tt}=g_{i}\cdots /(g_{j}\cdots)$.
Since $g_{tt}$ appears only in the form of $\ln g_{tt}$,
$g_{tt}$ becomes the solution of eq.(\ref{gtt})
if each $g_j$ satisfies
\begin{eqnarray}
\partial_\rho \{\rho \partial_\rho (\ln g_i) \}+
\partial_z \{\rho \partial_z (\ln g_i) \}=0.\label{mudiff}
\end{eqnarray}
In fact, if $g_i=\mu_i$ or $-\rho^2$,
eq.(\ref{mudiff}) holds.
Therefore, $g_{tt}$ which is a product of $\rho^2$ and $\mu_i$ 
satisfies eq.(\ref{gtt}).
We can construct $g_{\psi\psi}$ and $g_{\phi\phi}$ similarly.
Moreover, $\det {\bf g}= -\rho^2$ is achieved 
provided that the total number of each $\mu_k$ in the numerator 
in all components is equal to that in the denominator 
and that $\rho^2$ appears once in the numerator among
all components.
This means that ${\bf g}$ given in (\ref{gg}) satisfies 
eq.(\ref{diffg}) and $\det {\bf g}=-\rho^2$ 

We must also solve eq.(\ref{rhof}) and eq.(\ref{zf}).
Thanks to the diagonal form of the metric ${\bf g}$, the traces 
$\mbox{Tr}(U^2-V^2)$ and $\mbox{Tr}(UV)$ in these equations 
become the summations of the contributions from each component of ${\bf g}$.
In addition, since $f$ appears only in the form of $\ln f$ in the equations (\ref{rhof}) and (\ref{zf}), 
the left hand sides of these equations can be written as a summation of $\ln f_m$ 
with $f=f_1f_2 \cdots f_n$.
By this fact, we have only to solve for $f_m$ corresponding 
to each term of the right hand side of eq.(\ref{rhof}) and eq.(\ref{zf}).
First, we consider a component without $\rho^2$, 
which is written as
\begin{eqnarray}
\frac{\mu_{i_1}\cdots \mu_{i_m}}{\mu_{i'_1}\cdots \mu_{i'_n}}.
\end{eqnarray}
Then $(1/4 \rho) \mbox{Tr}(U^2-V^2)$ and $(1/2 \rho)\mbox{Tr}(UV)$ becomes
\begin{eqnarray}
&&\frac{1}{4\rho}\mbox{Tr}(U^2-V^2) \supset  
\frac{\rho}{4}\sum_{k}\left(\frac{(\partial_\rho \mu_{i_k})^2-(\partial_z \mu_{i_k})^2}{ \mu_{i_k}^2}  \right)
+\frac{\rho}{4}\sum_{k}\left(\frac{(\partial_\rho \mu_{i'_k})^2-(\partial_z \mu_{i'_k})^2}{ \mu_{i'_k}^2}  \right)
\nonumber\\
&&\qquad\qquad\qquad\qquad\qquad
+\frac{\rho}{2}\sum_{k\neq l}\left(
\frac{(\partial_\rho \mu_{i_k})(\partial_\rho \mu_{i_l})-(\partial_z \mu_{i_k})(\partial_z \mu_{i_l})}
{ \mu_{i_k} \mu_{i_l}}  \right)
+\frac{\rho}{2}\sum_{k \neq l}\left(
\frac{(\partial_\rho \mu_{i'_k})(\partial_\rho \mu_{i'_l})-(\partial_z \mu_{i'_k})(\partial_z \mu_{i'_l})}
{ \mu_{i'_k} \mu_{i'_l}}  \right)
\nonumber\\
&&\qquad\qquad\qquad\qquad\qquad\qquad\qquad\qquad\qquad\qquad\qquad\qquad\qquad
-\frac{\rho}{2}\sum_{k,l}\left(
\frac{(\partial_\rho \mu_{i_k})(\partial_\rho \mu_{i'_l})-(\partial_z \mu_{i_k})(\partial_z \mu_{i'_l})}
{ \mu_{i_k} \mu_{i'_l}}  \right),
\end{eqnarray}
and
\begin{eqnarray}
&&\frac{1}{2\rho}\mbox{Tr}(UV)\supset 
\frac{\rho}{2}\sum_{k}\left(\frac{\partial_\rho \mu_{i_k}\partial_z \mu_{i_k}}{ \mu_{i_k}^2}  \right)
+\frac{\rho}{2}\sum_{k}\left(\frac{\partial_\rho \mu_{i'_k}\partial_z \mu_{i'_k}}{ \mu_{i'_k}^2}  \right)
\nonumber\\
&&\qquad\qquad\qquad\qquad\qquad
+\frac{\rho}{2}\sum_{k\neq l}\left(
\frac{(\partial_\rho \mu_{i_k})(\partial_z \mu_{i_l})+(\partial_z \mu_{i_k})(\partial_\rho \mu_{i_l})}
{ \mu_{i_k} \mu_{i_l}}  \right)
+\frac{\rho}{2}\sum_{k\neq l}\left(
\frac{(\partial_\rho \mu_{i'_k})(\partial_z \mu_{i'_l})+(\partial_z \mu_{i'_k})(\partial_\rho \mu_{i'_l})}
{ \mu_{i'_k} \mu_{i'_l}}  \right)
\nonumber\\ 
&&\qquad\qquad\qquad\qquad\qquad\qquad\qquad\qquad\qquad\qquad\qquad\qquad\qquad
-\frac{\rho}{2}\sum_{k,l}\left(
\frac{(\partial_\rho \mu_{i_k})(\partial_z \mu_{i'_l})+(\partial_z \mu_{i_k})(\partial_\rho \mu_{i'_l})}
{ \mu_{i_k} \mu_{i'_l}}  \right).
\end{eqnarray}
As we explained above, we only have to solve the equations
\begin{eqnarray}
&&\partial_\rho \ln f_m = 
\frac{1}{2\rho}\left(
\frac{(\partial_\rho \mu_k)(\partial_\rho \mu_l)-(\partial_z \mu_k)(\partial_z \mu_l)}
{ \mu_k \mu_l}  \right),
\\
&&\partial_z \ln f_m = \frac{1}{2\rho} \left(
\frac{(\partial_\rho \mu_k)(\partial_z \mu_l)+(\partial_z \mu_k)(\partial_\rho \mu_l)}
{ \mu_k \mu_l}  \right).
\end{eqnarray}
A solution of these equations is given by
\begin{eqnarray}
f_m = \frac{\mu_k\mu_l}{\rho^2+\mu_k\mu_l}.
\end{eqnarray}
Next, we consider the component with $\rho^2$.
The difference from the case of the component without $\rho^2$ is that 
$(1/4 \rho) \mbox{Tr}(U^2-V^2)$ and $(1/2 \rho) \mbox{Tr}(UV)$, respectively, have extra terms
\begin{eqnarray}
\frac{1}{\rho}
+ \sum_{k}\frac{\partial_\rho \mu_{i_k}}{\mu_{i_k}}
-\sum_{k}\frac{\partial_\rho \mu_{i'_k}}{\mu_{i'_k}},\label{mumu}
\end{eqnarray}
and
\begin{eqnarray}
\sum_{k}\frac{\partial_z \mu_{i_k}}{\mu_{i_k}}
-\sum_{k}\frac{\partial_z \mu_{i'_k}}{\mu_{i'_k}}.\label{mumumu}
\end{eqnarray}

The first term of (\ref{mumu}) cancels $-1/\rho$ on the right hand side of eq.(\ref{rhof}).
$f_m$ corresponding to the the second and the third of  (\ref{mumu}) and (\ref{mumumu}) is $\mu_k$.
As a result,  remembering that $f$ can be written with the product of $f_m$,
$f$ is obtained as eq.(\ref{F}).

\end{document}